\documentclass{aa}
\usepackage{graphicx}
\def\deg{\ifmmode^{\circ}\;\else$^{\circ}\;$\fi} % overwrites \deg in LaTeX

\def\ni{\noindent}
\def\beb{\begin{thebibliography}{999}}
\def\eeb{\end{thebibliography}}
\def\bei{\begin{itemize}}
\def\eei{\end{itemize}}
\def\bef{\begin{figure}}
\def\eef{\end{figure}}
\def\ben{\begin{enumerate}}
\def\een{\end{enumerate}}
\def\beq{\begin{equation}}
\def\eeq{\end{equation}}
\def\ber{\begin{eqnarray}}
\def\eer{\end{eqnarray}}

\newcommand{\be}{\begin{equation}}
\newcommand{\ee}{\end{equation}}

\begin{document}

\title{Torque decay in the pulsar $(P,\dot{P})$ diagram}
\subtitle{Effects of crustal ohmic dissipation and alignment}

\author{T. M. Tauris\inst{1}
        \and
        S. Konar\inst{2}
        } 
\offprints{T. M. Tauris}
\institute{Nordic Institute for Theoretical Physics (NORDITA),
           Blegdamsvej 17, 2100 Copenhagen {\O}, Denmark\\
           \email{tauris@nordita.dk}
           \and
           Inter-University Centre for Astronomy and Astrophysics,
           Pune 4111007, India\\
           \email{sushan@iucaa.ernet.in}
}

\date{Received 30 January 2001 / Accepted 9 July 2001}

\abstract{
We investigate the evolution of pulsars in the $(P,\dot{P})$ diagram.
We first present analytical formulae to follow the evolution of a
pulsar using simple exponential
models for magnetic field decay and alignment.
We then compare these evolutionary tracks with detailed model
calculations using ohmic decay of crustal neutron star
magnetic fields. 
We find that, after an initial phase with a small braking index, $n$,
pulsars evolve with enhanced torque decay ($n\gg 3$) for about 1~Myr. 
The long term evolution depends on the impurity parameter of the crust.
%If $Q\simeq 0$ in older isolated pulsars we expect their
If impurities are negligible in older isolated pulsars we expect their
true age to be approximately equal to their observed
characteristic age, $\tau = P/(2\dot{P})$.
It is not possible from data to constrain model parameters of the neutron star crust.
}

\maketitle

\section{Introduction}
One of the key issues in pulsar physics is the question of torque
decay. It is important to determine to what extent the braking
torque decays in order to follow the dynamical evolution
and calculate the true age of pulsars. 
The time dependent behaviour of the torque is closely linked to
the important question of magnetic field decay and alignment of
the magnetic and rotational axes of a neutron star. 

In Fig.~1 we have plotted the rotational period, $P$, versus its derivative,
$\dot{P}$,
for 947 observed pulsars (F. Camilo, private communications).
The pulsars evolve to the right in this diagram as they loose
rotational energy.
To date, only four pulsars (see Lyne et al. 1996 and references therein) 
have a measured value of $\ddot{P}$ (and the so-called  braking index,
$n=\Omega \ddot{\Omega}/\dot{\Omega}^2$, where $\Omega=2\pi/P$)
which allows for a determination of the 
orientation of the evolutionary track in the $(P,\dot{P})$ diagram
for the given pulsar at its present age. However, these four
pulsars are all very young ($\tau \equiv P/2\dot{P} \simeq 2-15$ kyr) 
-- otherwise $\ddot{P}$ is too difficult to measure. Thus we cannot
learn much about long term torque decay from these observations.\\
Here we will first derive analytical evolutionary tracks by
considering simple $\vec{B}$-field decay and alignment. Both of these
effects will give rise to enhanced torque decay and deviation from evolution
along a straight line in the $(P,\dot{P})$ diagram. Thereafter we shall
present evolutionary tracks calculated from ohmic decay of crustal
neutron star magnetic fields. Such calculations of the neutron star
crust were introduced by Sang \& Chanmugam~(1987) and
Urpin \& Muslimov~(1992) who also included cooling models.

The millisecond pulsars ($P\la$~0.1~sec and
$\dot{P}\la10^{-17}$) are seen to form a separate population in the 
$(P,\dot{P})$ diagram shown in Fig.~1 -- presumably as a result of their binary origin
(Alpar et al. 1982; van den Heuvel 1984). 
In this paper we shall only concentrate on the evolution of isolated pulsars. 
For calculations of accretion-induced
magnetic field decay of a binary neutron star (the millisecond pulsar progenitor)
see e.g. Geppert \& Urpin~(1994), Urpin, Geppert \& Konenkov~(1997),
Konar \& Bhattacharya (1997; 1999).

We outline the dynamics of a rotating neutron star in Sect.~2.
In Sect.~3 we discuss a simple combined model with exponential
magnetic field decay and alignment. In this section we also
discuss the observed data and evidence for enhanced torque decay,
as well as the role of the inclination angle with respect to the
braking torque. A detailed model of the crustal physics of a cooling
neutron star is used in Sect.~4 to calculate more realistic
evolutionary tracks in the ($P,\dot{P}$) diagram.
Finally, the conclusions are briefly summarized in Sect.~5.

%--------------------------------------------------------------------------
\begin{figure}
    \centering
    \includegraphics[height=11.0cm,width=8.5cm]{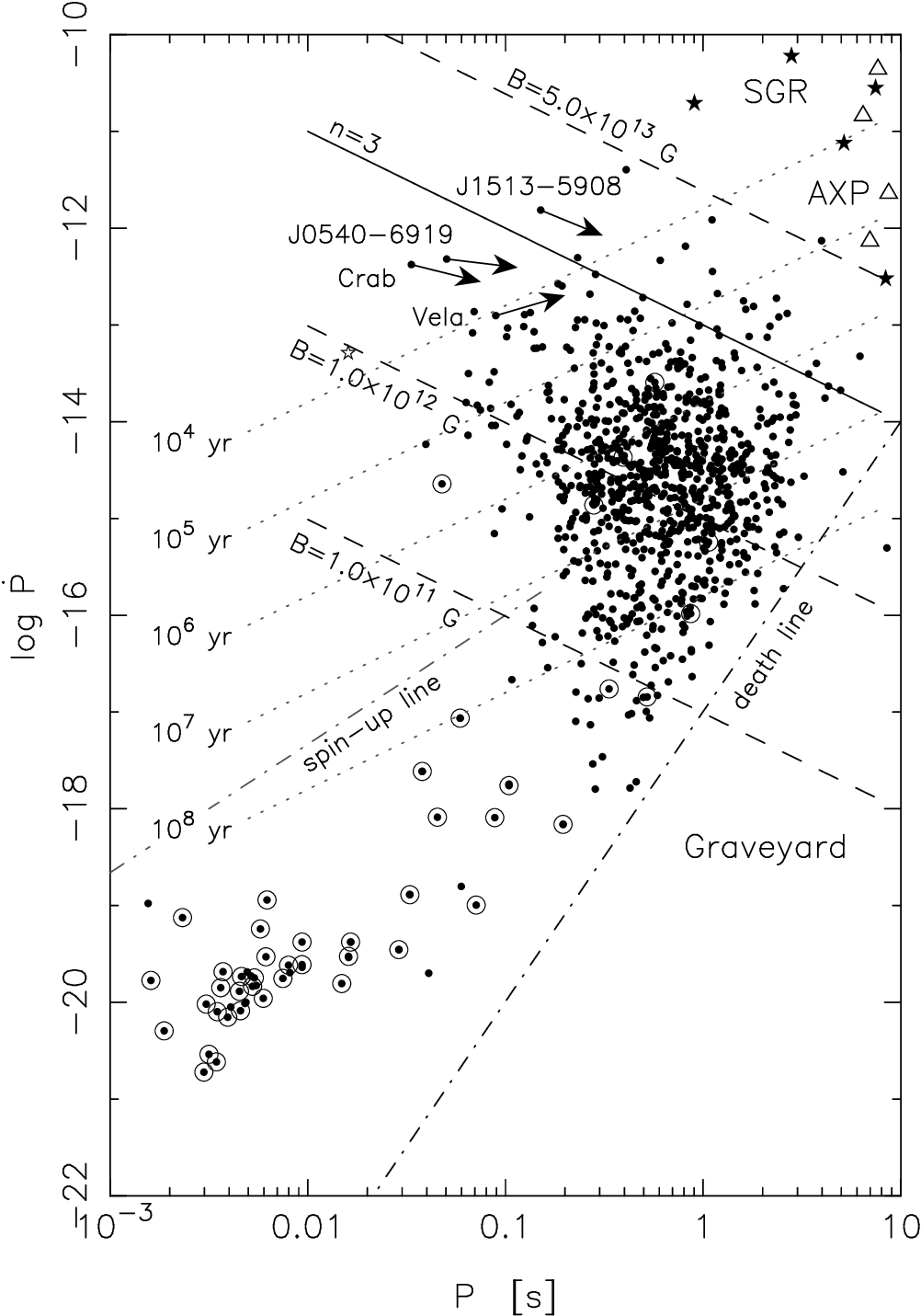}
 \caption{Distribution of 947 observed pulsars in the 
          $(P,\dot{P})$ diagram. Binary pulsars are marked with a circle.
          The four pulsars with measured braking
          index evolve along the direction of the arrows.
          The solid line is an evolutionary track with $n=3$. The dotted
          and dashed lines represent constant characteristic age and
          constant magnetic field, respectively. When radio pulsars
          cross the ``death line" their emission is terminated as a result
          of the low electrostatic potential drop across their polar cap.
          In the top right corner is plotted observations of
          soft $\gamma$-ray repeaters (SGR) and anomalous {X}-ray pulsars (AXP).
\label{fig1}}
\end{figure}
%--------------------------------------------------------------------------

\section{The rotating neutron star}
In the following section we shall
concentrate on the ``slow'' (non-recycled) pulsars whose evolution
is not ``polluted'' by interaction with a companion star.\\
Consider a neutron star with moment of inertia, $I$ and
angular frequency, $\Omega$. Its rotational
angular momentum and spin period is given by:
\begin{equation}
L=I\Omega \qquad P=2\pi/\Omega
\end{equation}
The rotational energy and its loss rate are given by:
\begin{equation}
  E_{\rm rot}=\frac{1}{2}I\Omega^2 \qquad \dot{E}_{\rm rot}=I\Omega\dot{\Omega}
\end{equation}
The braking torque acting on the neutron star is thus:
\begin{equation}
  N\equiv\frac{dL}{dt}=I\dot{\Omega}=\dot{E}_{\rm rot}/\Omega
\end{equation}
(if we assume that $I$ is constant in time).\\
For an isolated neutron star $\dot{E}_{\rm rot}<0$, and hence $N<0$.
The change in magnitude of the torque as a function of time is 
given by:
\begin{equation}
  \frac{d|N|}{dt} = - I \ddot{\Omega}
\end{equation}
The term ``torque decay'' thus refers to a decrease
in the {\it magnitude} of the torque.
Hence, the needed condition is:
\begin{equation}
  \frac{d|N|}{dt} < 0 \quad \Longleftrightarrow \quad n>0
\end{equation}
where we have introduced the braking index, $n$ defined as: 
\begin{equation}
  n \equiv \frac{\Omega\ddot{\Omega}}{\dot{\Omega}^2} =
           2 - \frac{P\ddot{P}}{\dot{P}^2}
\end{equation}
By introducing the characteristic age, $\tau\equiv P/(2\dot{P})$ we can
get an idea of how $|N|$ decreases with the age of a pulsar:
\begin{equation}
  N= I\dot{\Omega}= -\pi I\frac{1}{\tau P} \quad \rm{or} \quad 
  |N| \propto \frac{1}{\tau P}
\end{equation}

\subsection{Torque decay and braking index in the magnetic-dipole model}
In the magnetic-dipole model (Pacini 1967, 1968; Ostriker \& Gunn 1969)
the spin-down energy is carried away by magnetic dipole radiation:
\begin{equation}
  \dot{E}_{\rm rot}=I\Omega\dot{\Omega}=-\frac{2}{3c^3}|\ddot{\vec{m}}|^2=
          -\frac{2B^2R^6\Omega^4\sin^2\!\alpha}{3c^3}
\end{equation}
where $R$ is the radius of the neutron star,
$B$ is the strength of the
dipole component of the magnetic field at its equatorial surface,
$\vec{m}$ is the magnetic dipole moment and $\alpha$ is the
inclination of the magnetic axis with respect to the rotation axis.
Hence, in this model the braking torque 
(i.e. the radiation-reaction torque) transmitted to the neutron
star by the magnetic field is proportional to $B^2\,\sin^2\!\alpha$.
From the above equation one obtains:
\begin{equation}
P\dot{P}=\left( \frac{8\pi^2R^6}{3c^3I}\right)B^2\;\sin^2\!\alpha
        = \frac{1}{k^2}\, B^2 \sin^2\!\alpha 
\end{equation}
where we have assumed $k\equiv\sqrt{\frac{3}{8}c^3I}/\pi R^3$ is a constant.

\section{Evolution with $\vec{B}$-field decay and alignment}
Soon after the discovery of pulsars Ostriker \& Gunn (1969) suggested
that the magnetic field of a neutron star should decay exponentially
on a time-scale $\tau_D \sim 4$ Myr, because of
ohmic dissipation of the supporting currents.
The same authors also presented observational evidence in favour
of field decay (Gunn \& Ostriker 1970).
However, it was soon argued by others that the ohmic decay cannot be
important since the interior of a neutron star is likely to be
superconducting (Baym, Pethick \& Pines 1969).
Ever since the early days of pulsar astronomy this important 
question of magnetic field decay has continued to be rather controversial,
see e.g. Bhattacharya \& Srinivasan (1995) for a review on this matter.\\
Let us for a moment assume an exponential field decay\footnote{
This equation follows from Eq.~(A.8) in the appendix. From Eq.~(A.9)
one finds $\tau_{\rm D}\sim 4$ Myr, as suggested by 
Ostriker \& Gunn (1969), if $L\sim 10^6$ cm and $\sigma \sim 10^{22}$ s$^{-1}$.
However, this assumed value of the electrical conductivity, 
$\langle \,\sigma\rangle$ is too small.}:
\begin{equation}
  B(t)=B_0 \, e^{-t/\tau_D}
\end{equation}
where $\tau_D$ is the time-scale of decay and $B_0$ is the strength of
the surface dipole magnetic field at time $t_0=0$.

Over the years, many authors have argued for or against
observational evidence that the magnetic axis aligns with the rotation axis:
Proszynski~(1979), Candy \& Blair~(1983,1986), Lyne \& Manchester~(1988),
McKinnon~(1993), Gould~(1994) and Gil \& Han~(1996).
Beskin et al.~(1988) even
argue in favour of counter-alignment based on their magnetospheric theory.
Recently Pandey \& Prasad~(1996) and Tauris \& Manchester~(1998)
argue in favour of alignment.
It is therefore important to investigate the evolution of the
braking torque since this could be significantly influenced by the
alignment process.
The idea that the magnetic axis aligns with the rotation axis
was first analyzed analytically by Jones (1976). He suggested a simple
exponential decay of the inclination angle, $\alpha$ of the form:
\begin{equation}
  \sin\alpha(t)=\sin\alpha_0\,e^{-t/\tau_A}
\end{equation}
where $\tau_A$ is the alignment constant and $\alpha_0$ is the initial
inclination angle at the time $t_0=0$.

Inserting the two above expressions into Eq.~(9) and integrating yields the
period evolution of a pulsar in the general case including 
both exponential magnetic field decay and alignment:
\begin{equation}
  P^{2}(t) = \Pi_0 \, \tilde{\tau}_D \,
%              (1-e^{\displaystyle -2t/\tilde{\tau}_D})
               (1-e^{-2t/\tilde{\tau}_D})
           +\,P_0^2
\end{equation}
where $\Pi_0 \equiv P_0\dot{P}_0 = (B_0 \, \sin\alpha_0/k)^2$.
We have introduced an effective reduced time scale,
$\tilde{\tau}_D \equiv \frac{\tau_D \tau_A}{\tau_D+\tau_A}$
due to the symmetry of Eqs~(10) and (11).\\
Using Eqs~(3), (9), (12) and $\dot{\Omega}=-2\pi \dot{P}/P^2$ gives
the time dependence of the braking torque:
\begin{equation}
  N(t)=\frac{-2\pi I \,\Pi_0\; e^{-2t/\tilde{\tau}_D}}
            {\left[ \Pi_0\,\tilde{\tau}_D\,
                  (1-e^{-2t/\tilde{\tau}_D})
            +P_0^2 \right] ^{3/2}}
\end{equation}
A simple differentiation of Eq.~(8) yields:
\begin{eqnarray}
   \ddot{\Omega}(t)&=&-\frac{4R^6B(t)\dot{B}(t)\sin^2\!\alpha(t)}{3c^3 I}\;
                     \Omega^{3}(t) \nonumber \\
   && -\frac{4R^6B^{2}(t)\sin\alpha(t)\cos\alpha(t)\,\dot{\alpha}(t)}{3c^3I}\; 
       \Omega^{3}(t) \nonumber \\
   && -\frac{2R^6B^{2}(t)\sin^2\!\alpha(t)}{c^3I}\; 
       \Omega^{2}(t)\, \dot{\Omega}(t)
\end{eqnarray}
and hence we have an expression for the braking index:
\begin{eqnarray}
  n(t)=3&-&\frac{3c^3I\dot{B}(t)}
                {R^6B^{3}(t)\sin^2\!\alpha(t)\, \Omega^{2}(t)} \nonumber \\
     &-& \frac{3c^3I\cot\alpha(t)\,\dot{\alpha}(t)}
         {R^6B^{2}(t)\sin^2\!\alpha(t)\, \Omega^{2}(t)}
\end{eqnarray}
or simply:
\begin{equation}
  n(t)=3-F(t)\,\frac{\dot{S}}{S}
\end{equation}
where
\begin{equation}
  F(t)\equiv [3c^3I]/[R^6B^2(t)\sin^2\!\alpha(t)\,\Omega^2(t)] 
\end{equation}
and
\begin{equation}
 S=\left\{ \begin{array}{llll}
   B(t)\,\sin\alpha(t) & \mbox{\hspace{0.1cm}for B-decay and alignment} \\
   B(t)      & \mbox{\hspace{0.1cm}for B-decay only} \\
   \sin\alpha(t) & \mbox{\hspace{0.1cm}for alignment only} \\
   B_0\,\sin\alpha_0   & \mbox{\hspace{0.1cm}for const. B and no alignment}  \\
           \end{array}
         \right. 
\end{equation}
From the above equations we immediately obtain the allowed values of $n(t)$,
depending on the effect acting on it 
(cf. Table~9.1 of Manchester \& Taylor 1977): 
\begin{equation}
  n(t)\left\{ \begin{array}{llll}
   >3 & \mbox{\hspace{0.3cm}for} & \mbox{\hspace{0.1cm}$\dot{B}<0\;\;{\rm and/or}$} & \mbox{\hspace{0.1cm}$\dot{\alpha}<0$}\\ 
   <3 & \mbox{\hspace{0.3cm}for} & \mbox{\hspace{0.1cm}$\dot{\alpha}>0\,\;\;{\rm and}$} & \mbox{\hspace{0.1cm}$\dot{B}\not<0$}\\ 
   =3 & \mbox{\hspace{0.3cm}for} & \mbox{\hspace{0.1cm}$\dot{B}=0\;\;{\rm and}$}        & \mbox{\hspace{0.1cm}$\dot{\alpha}=0$}\\
               \end{array}
         \right.
\end{equation}
where $\dot{\alpha}>0$ corresponds to the case of counter-alignment.
            
\subsection{Constant magnetic field and no alignment}
A Taylor expansion of Eqs~(12), (13) and (15) 
for $\tau_D \rightarrow \infty$
or/and $\tau_A \rightarrow \infty$ yields the period, torque and braking index
in case of a constant magnetic field, no alignment or both. 
For example, if we assume no field decay and no alignment 
($\tau_D \rightarrow \infty$ and $\tau_A \rightarrow \infty$) we obtain:
\begin{equation}
  P^{2}(t) = 2\,\Pi_0 \;t +\,P_0^2
\end{equation}
\begin{equation}
  N(t)=\frac{-2\pi I \,\Pi_0}
            {(2\,\Pi_0\;t+P_0^2)^{3/2}}
\end{equation}
\begin{equation}
  n=3
\end{equation}

\subsection{Results}
In the top panels of
Fig.~2 we have plotted evolutionary tracks in the $(P,\dot{P})$ diagram
for various values of $\tilde{\tau}_D$ and $\Pi_0$. 
In the left panel we have plotted evolutionary tracks calculated without
enhanced torque decay. The right panel shows evolutionary tracks
including enhanced torque decay. For each initial starting point
three curves are plotted corresponding to an increasing time scale, $\tilde{\tau}_D$
(from left to right).
We have assumed $k=3.2 \times 10^{19}$ 
g$^{\frac{1}{2}}$ cm$^{-\frac{1}{2}}$ s$^{-\frac{3}{2}}$.
The small dots
represent data of observed pulsars.
It is seen that many evolutionary tracks fitting the data can be obtained from
different choices of $\Pi_0$ and $\tilde{\tau}_D$. 
The thin lines in the top left panel
are isochrones of 0, 1, 5, 10, 50, 100, 500 kyr, 1,
5, 10, 50 and 80 Myr, respectively.

To illustrate the difference between an evolution without
($\tilde{\tau}_D = \infty$) and with ($\tilde{\tau}_D$ = 10 Myr)
enhanced torque decay (due to magnetic field decay, a multipole or alignment)
in Fig.~2 we have also plotted 
$P$, $n$, $\tau$ and $N$, 
calculated from the equations derived earlier 
as a function of the true age, $t$.
%--------------------------------------------------------------------------
\begin{figure*}
   \centering
   \includegraphics[height=23.5cm,width=17.0cm]{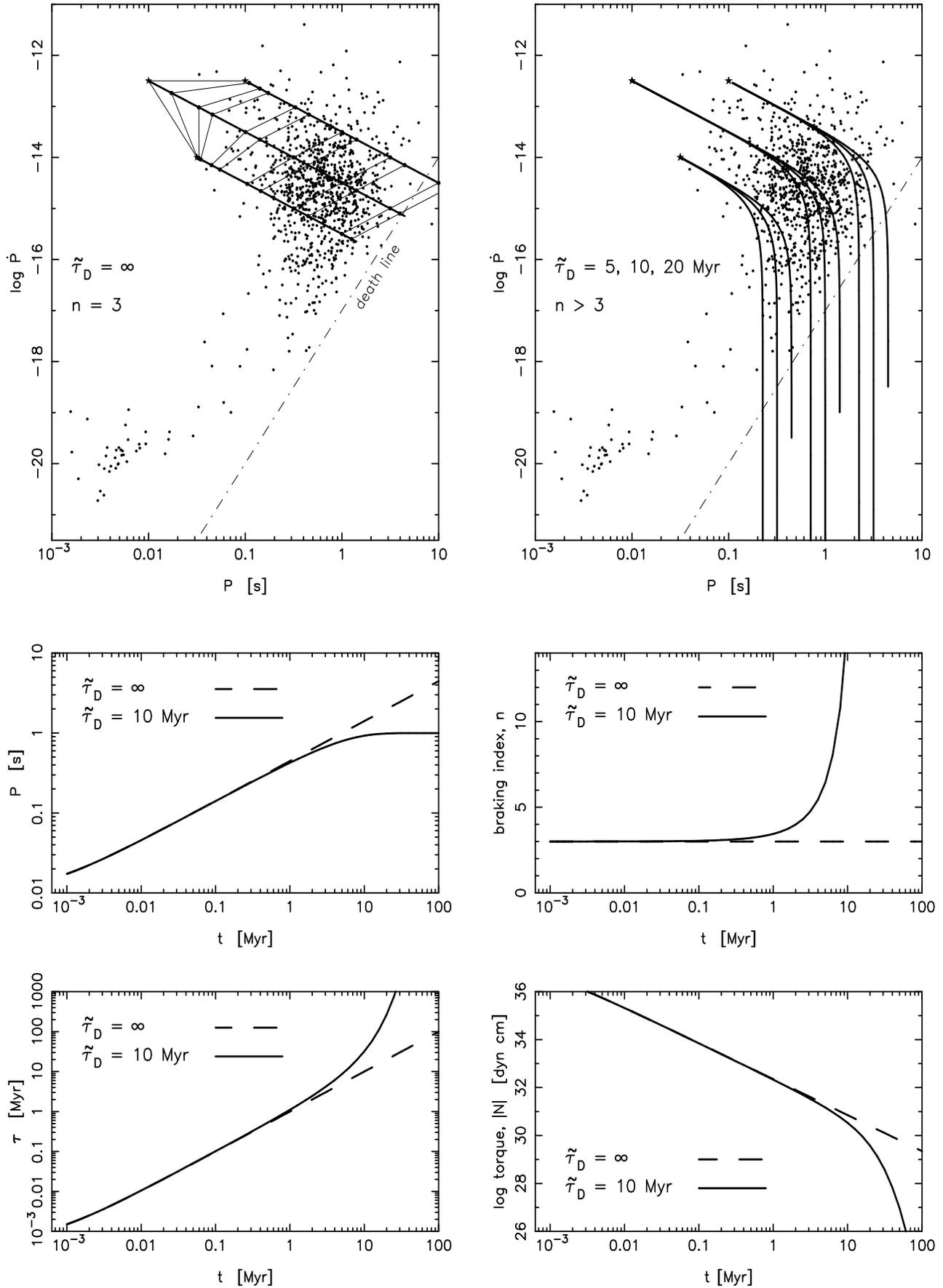}
 \caption{The top two panels show various evolutionary tracks in the
          $(P,\dot{P})$ diagram.
          The assumed time-scales for enhanced torque decay 
          are written in each panel. The bottom four panels show
          $P, n, \tau$ and $N$ as a function of true age, $t$ -- see text.
\label{fig2}}
\end{figure*}
%--------------------------------------------------------------------------
\begin{figure}
   \centering
   \includegraphics[height=6.5cm,width=8.5cm]{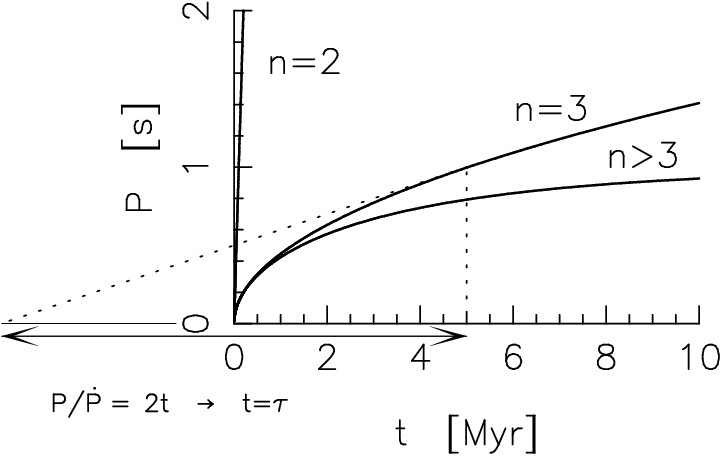}
 \caption{Dependence of the pulsar slow-down on the braking index, $n$. 
          The spin period, $P$ is plotted as a function of true age, $t$
          for $n=2$, $n=3$ and $n>3$ ($\tilde{\tau}_D = 10$ Myr). For $n=3$
          the slope of the curve is given by: $\dot{P}=P/(2t)$ and
          hence $t=\tau$ (where $P >\!\!> P_0=10.0$ ms).
          In Table~1 is given the relation between true age, $t$ and 
          characteristic age, $\tau$ for the different values of $n$.
\label{fig3}}
\end{figure}
%--------------------------------------------------------------------------
\begin{table}
         \caption[]{Values of true~age~($t$), characteristic~age~($\tau$),
                    spin period~($P$) and braking~index~$n(t)$ for three 
                    different models of evolution (birth parameters: $P_0=10$ ms and
                    $\log \dot{P}_0 = -12.5$).}
         \begin{center}
         \begin{tabular}{c|cc|cc|ccc}
          \cline{1-8}
          \noalign{\smallskip}
          & \multicolumn{2}{c}{$n=2$} & \multicolumn{2}{c}{$n=3$} & 
            \multicolumn{3}{c} {$n > 3\quad \tilde{\tau}_D=10$ Myr}\\
          \cline{2-8}
          \noalign{\smallskip}
          $t$ & $\tau$ & $P$ & $\tau$ & $P$ & $\tau$ & $P$ & n(t)\\
          Myr & Myr & s & Myr & s & Myr & s & \\
          \noalign{\smallskip}
          \cline{1-8}
          \noalign{\smallskip}
          0.01 & 0.01 & 0.10 & 0.01 & 0.05 & 0.01 & 0.05 & 3.00\\
          0.03 & 0.02 & 0.31 & 0.03 & 0.08 & 0.03 & 0.08 & 3.01\\
          0.10 & 0.05 & 0.99 & 0.10 & 0.14 & 0.10 & 0.14 & 3.04\\
          0.32 & 0.16 & 3.14 & 0.32 & 0.25 & 0.33 & 0.25 & 3.13\\
          1.00 & 0.49 & 9.94 & 1.00 & 0.44 & 1.10 & 0.42 & 3.44\\
          3.16 & 1.57 & 31.4 & 3.16 & 0.79 & 4.39 & 0.68 & 4.77\\
          10.0 & 4.98 & 99.4 & 10.0 & 1.41 & 31.8 & 0.93 & 15.8\\
          31.6 & 15.7 & 314  & 31.6 & 2.51 & 2774 & 1.00 & 1118\\
          100  & 49.8 & 994  & 100  & 4.46 & $>10^{9}$ & 1.00 & $>10^{8}$\\
          \cline{1-8}
         \end{tabular}
         \end{center}
\end{table}

%--------------------------------------------------------------------------

The evolution with enhanced torque decay (solid line, $n>3$) is clearly seen to
result in a strong decay of the braking torque for $t>\tilde{\tau}_D$ and hence 
a strong decrease of $\dot{P}$ ($\dot{P} \rightarrow 0$). 
Therefore a constant value of $P$ is approached
as the radio emission process becomes inactive after 10--50~Myr (while
the pulsar crosses the so-called death line).
The strong decrease in $\dot{P}$ has an important influence on the braking index
which accordingly increases sharply when $t \rightarrow \tilde{\tau}_D$ 
(see Eq.~6, $\ddot{P} < 0$). This is shown in the central right panel.

If pulsars evolve without magnetic field decay or alignment (dashed line, $n=3$)
they will attain much larger ages before they cross the death line.
For these pulsars the true age, $t$ will always be equal to 
the observed characteristic age, $\tau = P/2\dot{P}$ ($P \gg P_0$)
-- whereas if enhanced torque decay is important ($n>3$) the true age of
pulsars will be smaller than $\tau$ (in this case by more than an order of
magnitude for $\tau > 20$ Myr).\\ 
In Fig.~3 we have demonstrated this effect graphically.
The tangent to the $n=3$ curve at a given age, $t$
will always intersect the {x}-axis ($P=0$) at $t_{x}=-t$. Thus the slope of this tangent
is $\dot{P}(t) = P(t)/2t$, or equivalently $t=P(t)/2\dot{P}(t)$ which is equal
by definition to the characteristic age, $\tau$. The tangent to the $n>3$ curve
is seen to be less steep and therefore $\dot{P}(t) < P(t)/2t$, or $t<\tau$.
Finally, in the case of relaxed torque decay ($0<n<3$, e.g. radial deformation
of $\vec{B}$ or a decreasing moment of inertia in young pulsars) we have $t>\tau$.

For population statistics and evolution of pulsars it is therefore essential
to know whether or not enhanced torque decay occurs in order to determine
the true ages of the observed pulsars.\\
It should be noted, that strictly speaking the braking index, $n$ is defined
from the power-law: $\dot{\Omega} \propto -\Omega ^n$. However,
in general $n$ is a function of time: $\dot{\Omega}=\Lambda(t)\,\Omega^{n(t)}$
and since there is no analytical solution to this power-law equation,
we have retained the definition of $n$ as given in Eq.~(6).

\subsection{Observational evidence on torque decay ?}
Many attempts have been made to find observational evidence for
magnetic field decay. The key problem (as outlined above) is that
we do not know the true age of the pulsars observed. Furthermore, many
parameters connected to pulsars (e.g. $\tau$ and $B$) are given as
functions of the two fundamental observed parameters: $P$ and $\dot{P}$.
There are many different ways to illustrate the evolution of pulsars.
Here, we introduce another variant of the $(P,\dot{P})$ diagram
and plot the rotational energy of a pulsar as a function 
of its characteristic age. This is shown in Fig.~4. The dashed lines
represent evolution with $n=3$ (constant $B$-field and 
magnetic inclination angle) and the solid lines are calculated with enhanced
torque decay ($n>3, \tilde{\tau}_D = 10$ Myr). In each case we considered
two different sets of initial birth parameters for the pulsar:
($P_0=10.0$ ms, $\log \dot{P} = -13.5 \Rightarrow B_0=5.69\times 10^{11}$ G) and
($P_0=31.6$ ms, $\log \dot{P} = -12.0 \Rightarrow B_0=5.69\times 10^{12}$ G).
If pulsars evolve without enhanced torque decay they should move
along the dashed lines. If enhanced torque decay is important their
evolution should bend off along the solid lines.

We have shaded two regions in the diagram ({A} and {B}) where there is
a deficit of pulsars. If pulsars evolve with $n=3$, the region~{A}
should feed pulsars into the relatively dense populated area near the
arrowhead. Similarly, region~{B} should be the endpoint of pulsars
coming from the populated area along its arrow base.
So why are there so few pulsars in region~{A} and {B} ?
A simple reason for the few pulsars in region~{A} could be due to
the logarithmic scale for the characteristic age. Hence, the pulsars
accumulate with time near the arrowhead before crossing the death line.
However, if this is the correct explanation it seems strange why there
should not be a similar effect below this region, e.g. between and
parallel to the dashed lines. Furthermore, the deficit of pulsars in
region~{B} is even more strange in this respect (since pulsars are
then supposed to be accumulated in this region), unless the naively
drawn death line in reality bends into this region.
In both cases we notice that a natural explanation, for the deficit
of pulsars in the shaded regions, is the effect of enhanced torque decay
and subsequent bending of the evolutionary tracks.
We will let the reader draw his own conclusion.
%--------------------------------------------------------------------------
\begin{figure*}
   \centering
   \includegraphics[height=13.0cm,width=17.0cm]{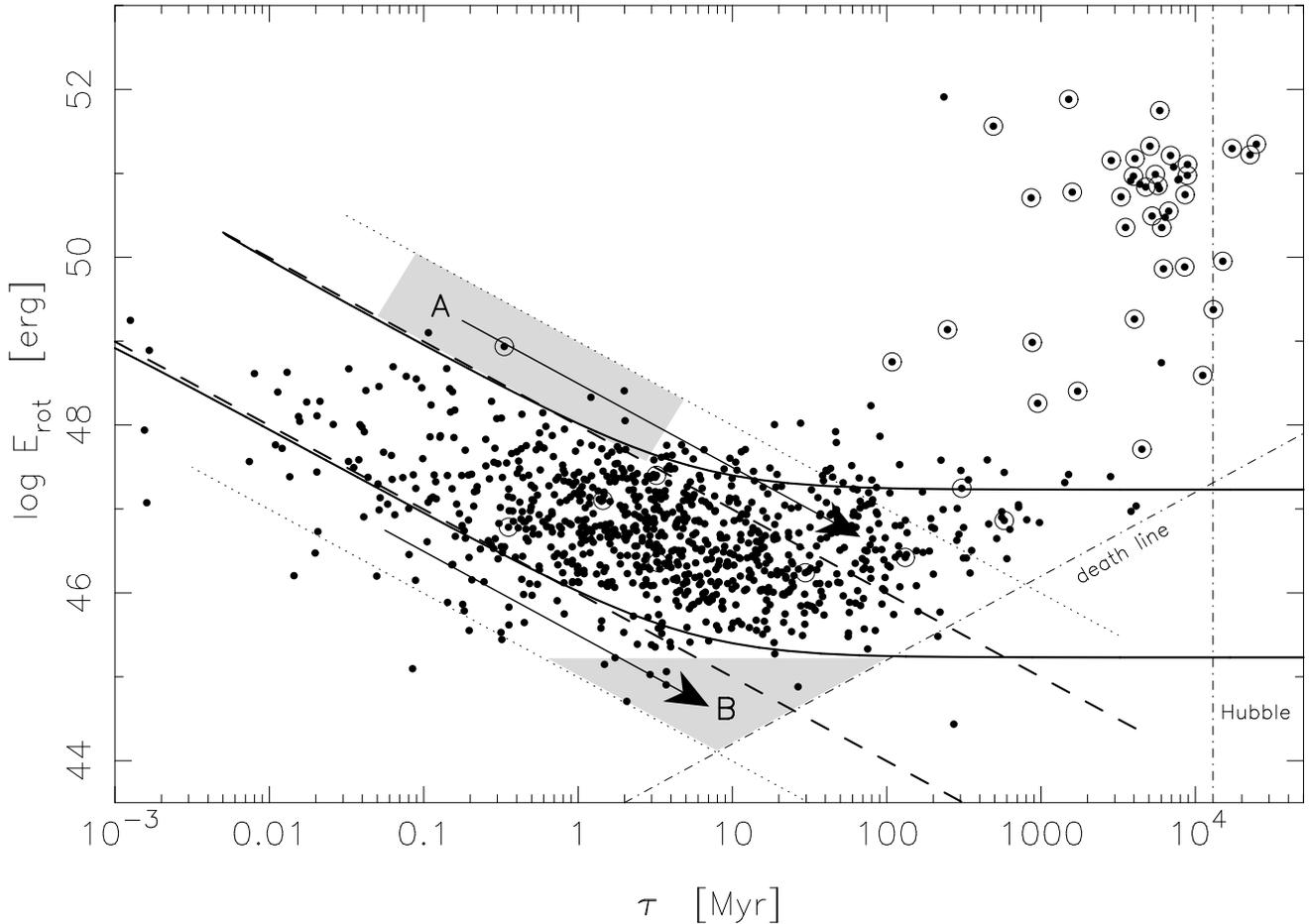}
 \caption{The rotational energy of observed pulsars plotted
as a function of their characteristic age. The solid and dashed 
lines are evolutionary tracks with and without enhanced torque decay,
respectively. We assumed a
pulsar moment of inertia, $I=10^{45}$~g~cm$^2$. The millisecond
pulsar population is seen in the upper right corner.
See text for discussion.
\label{fig4}}
\end{figure*}
%--------------------------------------------------------------------------

%--------------------------------------------------------------------------
\begin{figure}
   \centering
   \includegraphics[height=10.0cm,width=8.5cm]{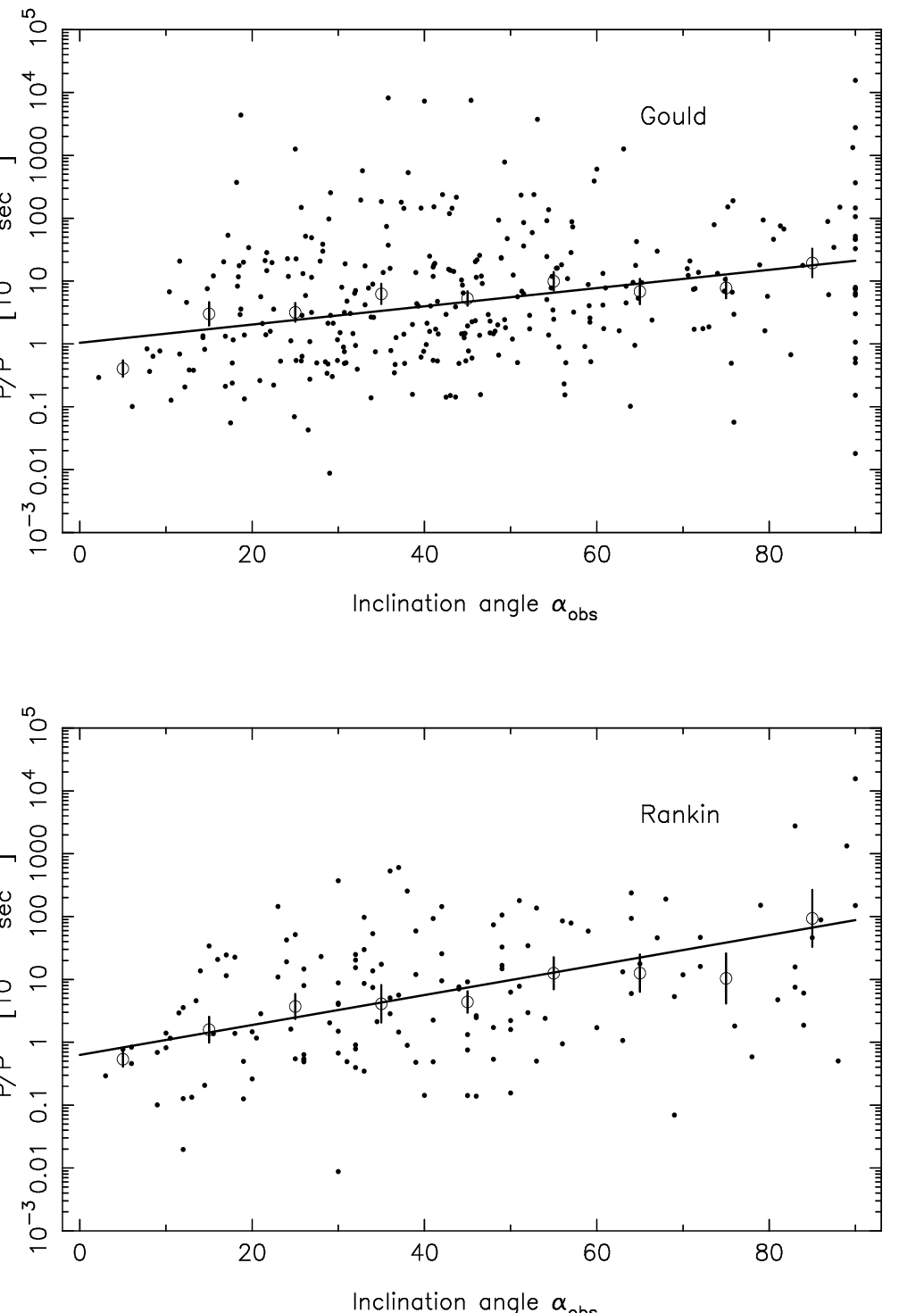}
 \caption{A plot of $\dot{P}/P^2$ vs. the observed inclination angle,
          $\alpha_{\rm obs}$. The top panel contains 316 pulsars
          obtained from Gould (1994) and the bottom panel shows 149
          pulsars from Rankin (1993) -- see Tauris \& Manchester (1998)
          for details.
          The braking torque, $N\propto \dot{P}/P^2$ is seen to be a
          slightly increasing function of $\alpha_{\rm obs}$.
          Standard error bars of the binned mean values are plotted.
\label{fig5}}
\end{figure}
%--------------------------------------------------------------------------
\subsubsection{Torque decay and the magnetic inclination angle}
In the Goldreich-Julian (1969) model of a non-vacuum magnetosphere of a
pulsar, the slow-down of rotation is caused by outflow of plasma:
\begin{equation}
  \dot{E}_{\rm rot} = - \left( \frac {B_{\rm L}}{8\pi} \right) ^2
                        4\pi R_{\rm L}^2\,c
\end{equation}
where $R_{\rm L}=c/\Omega$ is the radius of the light cylinder and
$B_{\rm L}$ is the magnetic field strength at this location --
for a dipole field: $B_{\rm L} \propto 1/r^3$ or $B_{\rm L}=B\,(R/R_{\rm L})^3$. 
This expression for
$\dot{E}_{\rm rot}$ is of the same magnitude as in Eq.~(8), but here
the braking torque ($N=\dot{E}_{\rm rot}/\Omega$) is independent of $\alpha$.
The dynamics of a rotating neutron star in this case is then similar to that
of the magnetic-dipole model for a constant inclination angle.
In Fig.~5 we have  plotted the braking torque, $N\propto \dot{P}/P^2$ vs.
the observed inclination angle, $\alpha_{\rm obs}$ for a large
number of pulsars. The derived, observed magnetic inclination
angles were taken from the work by Tauris \& Manchester~(1998)
and references therein.
Though we cannot rule out the possibility that $N$ is independent
of $\alpha$ (a $\chi^2$-test yields a $\sim 15$ \% probability 
for constancy) we notice that $N$ seems to be a slightly increasing 
function of $\alpha$ (although the scatter is large)\footnote{
This is in contrast to Bhattacharya (1989) who argued that
the spin-down torque is independent of $\alpha$. 
However, he considered explicitly the dipole model and
also assumed a constant magnetic field strength. This is why he
plotted $P\dot{P}$, rather than
$N=I\dot{\Omega} \propto\dot{P}/P^2$, as a function of $\alpha$.}.
However, we must also bear in mind that $\alpha$ itself is only a 
marginally decreasing function of characteristic age
(see Tauris \& Manchester~1998). Hence, based on observations
the effect of $\alpha$ on $N$ is probably not a major concern.

We have now demonstrated how simple analytic models can account for
the distribution of observed pulsars in the $(P,\dot{P})$ diagram,
although the behaviour of the evolutionary tracks is still an
unsettled question. The models we have used so far are based on crude
approximations (e.g. the exponential decay of the $B$-field).\\
In the following section we will apply more realistic physics to our
calculations and see how this will affect the calculated evolutionary tracks. 

\section{Ohmic decay of crustal neutron star $\vec{B}$-fields}
The physics and relevant equations of crustal $B$-field 
decay is given in the appendix.
We refer to Konar \& Bhattacharya~(1997; 1999) and Konar~(1997)
and references therein for further details on the computations.
Here we shall briefly summarize the assumptions necessary for 
the calculations we have performed in this paper\footnote{
For an alternative scenario of neutron star magnetic field evolution
caused by core field changes and crust movements, see e.g. Ruderman,
Zhu \& Chen~(1998).}.

We assume that the magnetic field has been generated in the outer
crust by some unspecified mechanism, e.g. by thermomagnetic effects
(Blandford, Applegate \& Hernquist 1983), during or shortly after
the neutron star is formed. 
For the distribution of the initial underlying toroidal current
we have chosen to use the configuration by Konar \& Bhattacharya~(1997).\\
The equation of state by Wiringa, Fiks \& Fabrocini~(1988) is
matched to Negele \& Vautherin~(1973) and Baym, Pethick \& Sutherland~(1971)
in order to construct a neutron star with a mass of $1.3\,\mathrm{M}_{\odot}$
($\mathrm{R}=11.1$ km).\\
The electrical conductivity in the neutron star crust is mainly a function of
mass density and temperature. The conductive properties of the solid crust are
determined by scattering of electrons on phonons and lattice impurities
(Yakovlev \& Urpin~1980 and Itoh~et~al.~1993). In the outer thin liquid
layer near the surface it is determined by scattering of electrons on ions
(e.g. Urpin \& Muslimov~1992).\\
We used the models by van~Riper~(1991a, 1991b) for standard cooling
of a neutron star (modified URCA process in the core and neutrino pair
bremsstrahlung in the crust). We adopted the profile of
Gudmundsson, Pethick \& Epstein 1982) to estimate the temperature
in the outer/surface regions ($\rho < 10^{10}$ g~cm$^{-3}$)
above the approximately isothermal inner crust/core region.

We have performed calculations of the ohmic decay of the crustal
magnetic field of an isolated pulsar for an age up to 100~Myr. 
In Fig.~6 we have plotted the evolution in the $(P,\dot{P})$ diagram.
Here we assumed that the
pulsar is born with an initial period, $P_0=10$ ms and
initial period derivative, $\dot{P}_0=10^{-11}$, corresponding
to an initial surface $B$-field strength of $B_0=10^{13}$ Gauss.
The three different main curves are the results of calculations
using different initial locations (densities) of the current 
configuration in the crust. The curves from top to bottom
correspond to $\log \rho_0=$ 13, 12 and 11 (g cm$^{-3}$), respectively.
The branching of the curves near their endpoints of evolution
is caused by different assumed values of the impurity parameter:
\begin{equation}
  Q \equiv \frac{1}{n_{\rm total}} \sum_i n_i (Z-Z_i)^2
\end{equation}
where $n_{\rm total}$ is the total ion density, $n_i$ is the density
of impurity species $i$ with charge $Z_i$, and $Z$ is the ionic
charge in the pure background lattice (Yakovlev \& Urpin 1980).
The four branches correspond to: $Q=$ 0, 0.01, 0.05 and 0.10
(top to bottom, respectively).
As the neutron stars cools sufficiently, increasing values of
$Q$ will decrease the electrical conductivity of the crust.
In Fig.~7 we demonstrate how a 
single model with fixed parameters
($\rho _0 = 10^{12}$~g~cm$^{-3}$ and $Q=0$) can cover the entire population
of observed non-recycled pulsars by choosing different initial values
of $\dot{P}_0$ (here corresponding to $\log B_0=$ 13.5, 13.0 and 12.5, 
keeping $P_0=10$ ms fixed).

The evolutionary tracks obtained from our calculations of
neutron star crustal field decay are also seen to cover
the observed distribution of pulsars in the ($P,\dot{P}$) diagram
quite well. However, from Figs.~(6) and (7) it is also clear that the
location of each track depends on the underlying parameters of the
applied model. It is impossible to constrain e.g. $\rho_0$ and $Q$
from the observed data set. Whereas a spread in the initial
magnetic field strengths, $B_0$ can be expected, other parameters,
like the impurity parameter, $Q$, should be more or less equal
for all pulsars.
%--------------------------------------------------------------------------
\begin{figure}
    \centering
    \includegraphics[height=9.5cm,width=7.0cm]{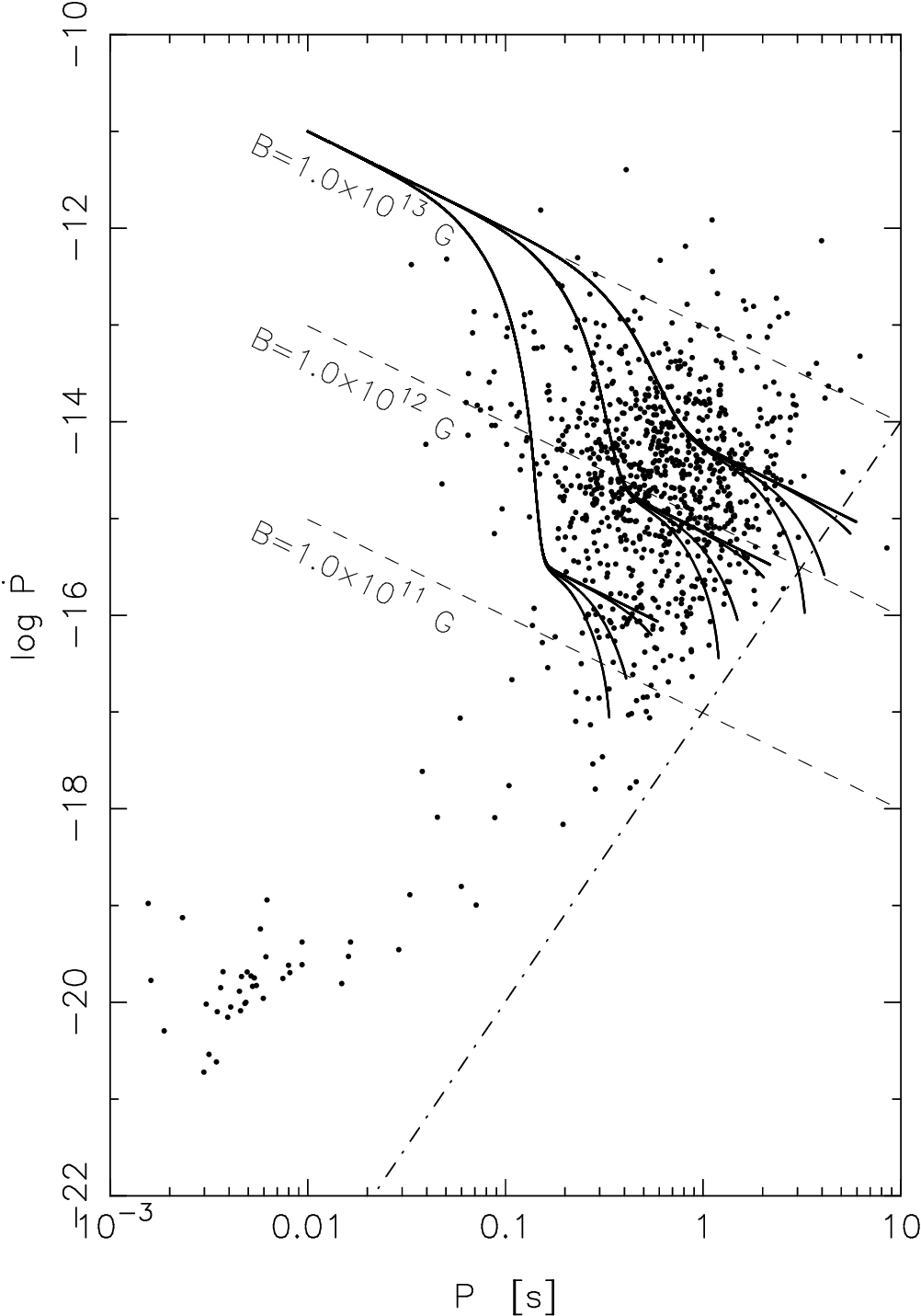}
\caption{Evolutionary tracks in the $(P,\dot{P})$ diagram
         assuming $P_0=10$ ms and $\dot{P}_0=10^{-11}$
         ($B_0=10^{13}$ G)
         calculated from $\rho_0=10^{11}-10^{13}$ g~cm$^{-3}$
         and $Q=0-0.10$ -- see text. Each track was
         followed for 100 Myr. 
         The small dots represent data.
\label{fig6}}
\end{figure}
%--------------------------------------------------------------------------
%--------------------------------------------------------------------------
\begin{figure}
    \centering
    \includegraphics[height=9.5cm,width=7.0cm]{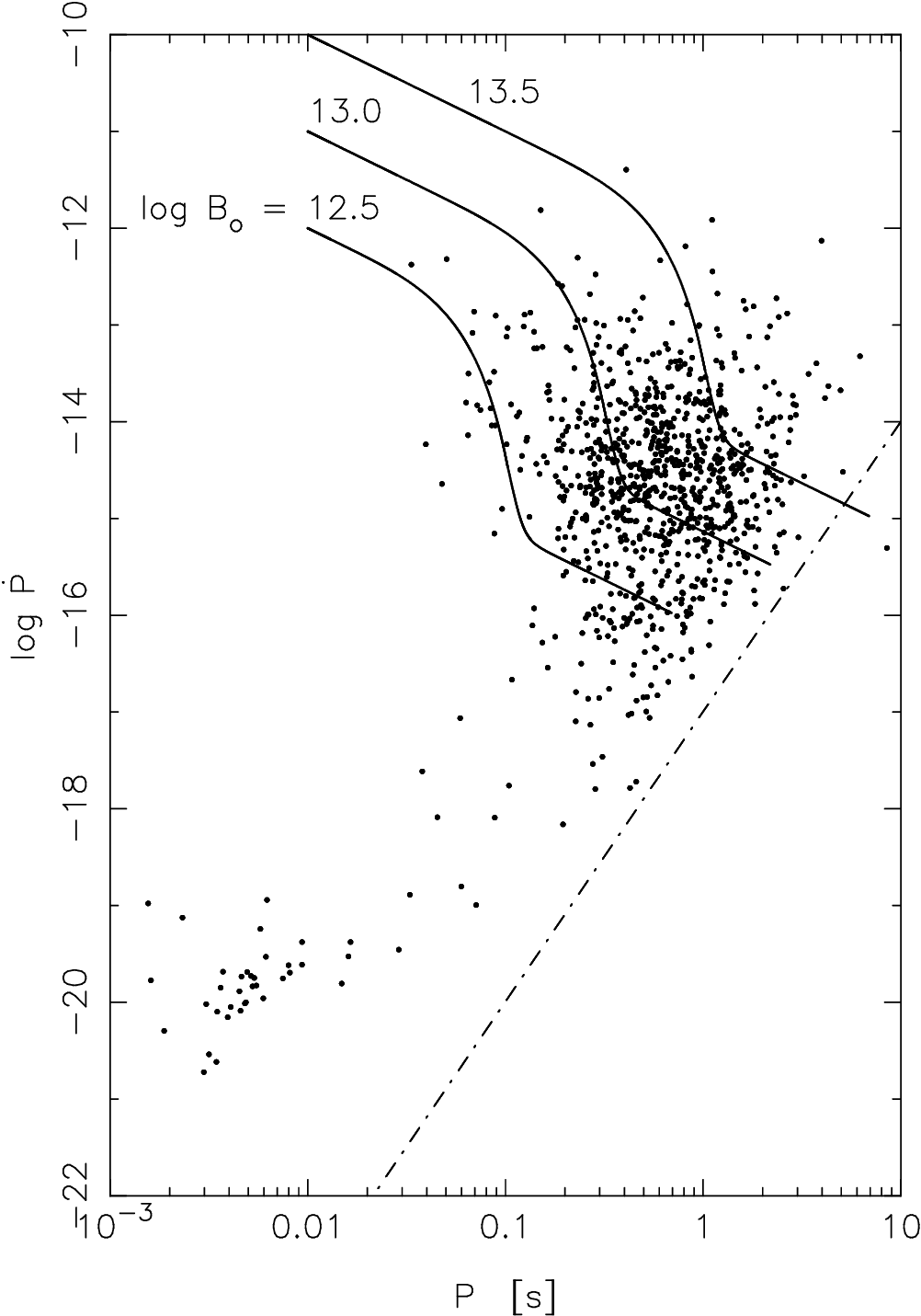}
 \caption{Evolutionary tracks calculated for 
          $\rho_0=10^{12}$ g cm$^{-3}$ and $Q=0$ at different
          initial $B$-fields for $P_0=10$ ms.
\label{fig7}}
\end{figure}
%--------------------------------------------------------------------------
%--------------------------------------------------------------------------
\begin{figure}
    \centering
    \includegraphics[height=6.5cm,width=7.0cm]{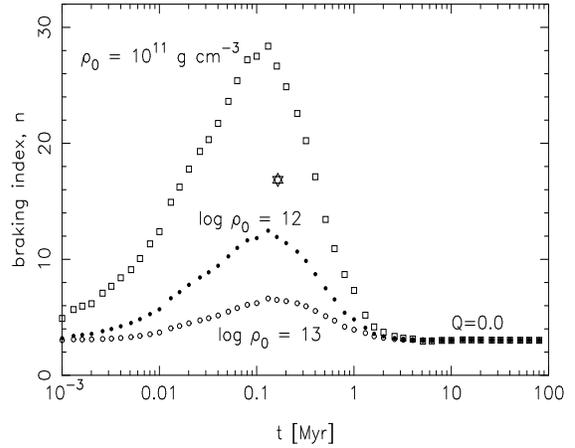}
 \caption{The braking index as a function of true age calculated for 
          a pulsar with crustal magnetic field decay.
          We assumed $B_0=10^{13} G$ ($P_0=10$ ms), $Q=0.0$ and
          $\log \rho_0$=11, 12 and 13 (g cm$^{-3}$). The star indicates 
          an average value obtained from Johnston \& Galloway (1999)
          -- see text for discussion.
\label{fig8}}
\end{figure}
%--------------------------------------------------------------------------

The calculated evolutionary tracks, of pulsars with ohmic crustal magnetic
field decay, are seen to initially evolve along straight ($n=3$) lines
in the $(P,\dot{P})$ diagram (Figs.~6 and 7), then bending down ($n>3$) for a while
and returning again to an evolution along a straight ($n=3$) line, if $Q=0$.
This is also shown in Fig.~(8) where we have plotted the braking index, $n$ 
as a function of true age, $t$. 
The evolution of $n$ as a function of $\log t$ can roughly be described to be
Gaussian-like for the first 10~Myr. Initially, $n=3$ (by assumption), then follows a short 
interval with $3<n<30$ before it settles down again and evolves with $n\approx 3$.
The more exact evolution in this temporary interval ($0.01 \la t \la 1$ Myr) 
depends on the initial location (depth) of the supporting currents, $\rho_0$ 
and the details of the cooling process (exotic vs. standard cooling).
The reason for the flattening of the evolutionary track (returning to $n=3$)
at ages $t> 1$ Myr, is due to a saturation of the decay of the $B$-field. 
The current distribution, which is responsible for the $B$-field,
migrates inward as a result of diffusion and enters the highly conducting parts 
of the neutron star. In that region $\tau _{\rm Ohmic} > \tau _{\rm Hubble}$ and hence
the magnetic field will essentially be stable forever -- it freezes out at
a residual value (Konar~1997).\\
Since the {x}-axis in Fig.~8 is logarithmic, a pulsar with $Q\simeq0$ spends the majority of its life
as an active pulsar evolving with $n\approx 3$. Hence, near the end of the
evolutionary tracks ($t \approx 100$ Myr) the characteristic age, $\tau$ is only
$\sim 3 \%$ larger than the true age, $t$.\\
The star in the centre of the figure is the mean value of $n$ for the eight pulsars
with the smallest error bars in Table~1 of Johnston \& Galloway~(1999).
Note, that they quote characteristic pulsar ages, $\tau$ and we have not corrected
for the unknown reduction in the true ages, $t$. Although there are large uncertainties
involved, it is interesting to notice that our calculations predict values of 
the braking index, $n$ within the range of their estimates.
%--------------------------------------------------------------------------
\begin{figure}
    \centering
    \includegraphics[height=6.5cm,width=7.0cm]{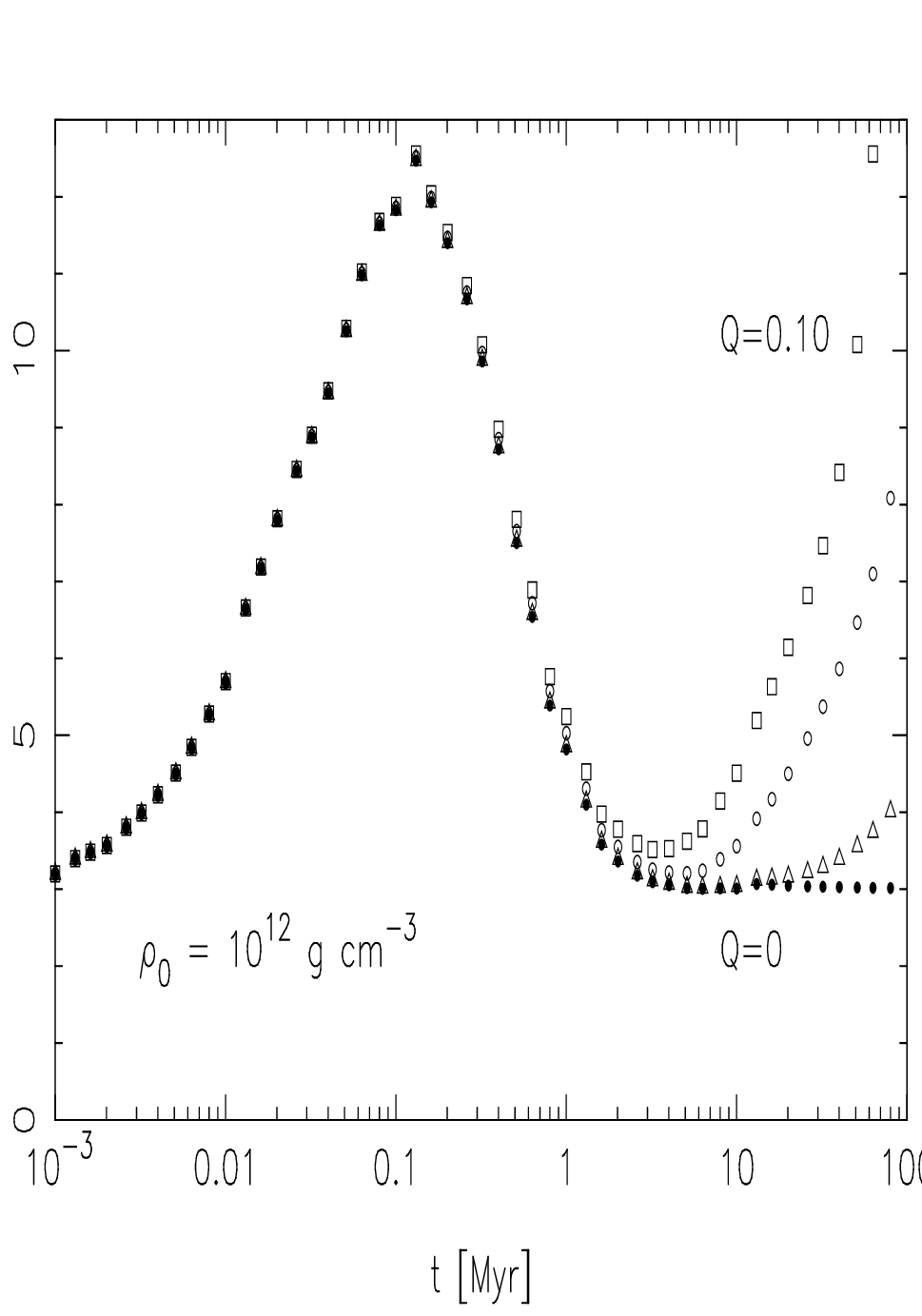}
 \caption{The braking index as a function of true age calculated for 
          a pulsar with crustal magnetic field decay.
          We assumed $B_0=10^{13} G$ ($P_0=10$ ms), $\rho_0=10^{12}$ g cm$^{-3}$ 
          and $Q=0$, 0.01, 0.05 and 0.10. See text for discussion.
\label{fig9}}
\end{figure}
%--------------------------------------------------------------------------

It has recently been suggested that the impurity parameter is probably
quite large in a heated crust of an accreting binary neutron star as a
result of nuclear processes (Brown \& Bildsten 1998).
However, it is uncertain whether or not $Q$ is important for isolated pulsars.
If there are significant impurities ($Q>0$) in the crustal crystalline lattice
of isolated pulsars, the braking index will increase again after $\sim 10$~Myr.
This is demonstrated in Fig.~9 (see also the final branching of the curves in
Fig.~6). Hence, a determination of the braking index for such old pulsars
could in principle yield the value of $Q$. Unfortunatly such direct measurements
of $n$ for old pulsars are not possible (see e.g. Johnston \& Galloway 1999).

\section{Conclusion}
The effects of long term torque decay are presently not known very well.
The only pulsars with a measurable braking index are quite young.
If more young pulsars turn out to have $2 \la n < 3$ it seems clear 
that $n$ must be increasing with age. Otherwise, their evolution will
not be consistent with the bulk of observed pulsars with ages $1-10$ Myr.
As we have argued in this paper, enhanced torque decay ($n>3$)
seems to play an important role for pulsars.
The exponential models for $B$-field decay and alignment are
obviously naive. However, they give a better description of the
observed data than those models with $n=3$.

Using a detailed model of the crustal physics of a neutron star, we have
demonstrated that pulsars may evolve for a limited interval of time
(from $0.01-1$ Myr) with enhanced torque decay ($n\gg 3$).
This can explain the distribution of
observed pulsars in the ($P,\dot{P}$) diagram quite well.
The subsequent evolution depends on the impurities in the crustal
crystalline lattice of old isolated pulsars. If $Q \simeq0$ the older 
pulsars ($t>10$ Myr) returns to long term $n\approx 3$ evolution and
future observations of the temperature of nearby,
relatively old pulsars (e.g. PSR~J0108--1431, Tauris~et~al.~1994)
may be able to verify or reject our hypothesis that 
$t\approx \tau$ for older pulsars.

Further investigations of the present pulsar data in the ($P,\dot{P}$) diagram
and statistical population analyses are encouraged in order to help verify 
our finding of enhanced torque decay at some stage during the pulsar evolution.

\begin{acknowledgements}
      We would like to thank Dipankar Bhattacharya
      and the Raman Research Institute (where some of
      this work was performed)
      for very nice and warm hospitality in November 1999.
      We also thank the referee, Ulrich Geppert, for pointing
      out a long term evolutionary dependence on $Q$.
      T.M.T. acknowledges the receipt of a NORDITA fellowship.
\end{acknowledgements}

\appendix
\section{Physics of the $\vec{B}$-field decay}
\label{smn-phys}
\ni The induction equation is given by (Jackson, 1975):
\beq
\frac{\partial \vec B}{\partial t} = - \frac{c^2}{4 \pi} \vec \nabla \times
(\frac{1}{\sigma} \times \vec \nabla \times \vec B)  
 + \vec \nabla \times (\vec V \times \vec B)     \label{emhd}
\eeq
where $\vec V$ is the velocity of material movement and $\sigma$ is the electrical conductivity of 
the medium. We choose the vector potential to be of the form ${\vec A} = (0, 0, A_{\phi})$
to ensure a poloidal geometry for $\vec B = \nabla \times \vec A$.
In particular:
\beq
A_{\phi} = \frac{g(r,t) \sin \theta}{r}
\eeq
where $g(r,t)$ is the Stokes' stream function.
In the lowest order of multipole, the dipolar form of the $\vec{B}$-field is:
\ber
\vec B(r, \theta) &=& \vec \nabla \times (\frac{g(r,t) \sin \theta}{r} \; \hat \phi) \nonumber \\
&=&  \frac{2 \cos \theta g(r,t)}{r^2} {\hat r} 
- \frac{\sin \theta}{r}\frac{\partial g(r,t)}{\partial r} {\hat \theta} \label{eBrtheta}
\eer
The magnitude of the field is then given by: 
\beq
B(r, \theta) = {\left[\frac{4 \cos^2 \theta g^2(r,t)}{r^4}
+ \frac{\sin^2 \theta}{r^2}(\frac{\partial g(r,t)}{\partial r})^2 \right]}^{1/2}
\eeq
Therefore, at the pole ($\theta = 0$, $r =  R$) the magnitude is:
\beq
B(R, 0) = \frac{2 g(R,t)}{R^2} \label{eBmag}
\eeq
which is simply proportional to the value of the Stokes' function there. The underlying current 
distribution corresponding to the above field is given by: 
\ber
\vec J &=& \frac{c}{4 \pi} \nabla \times \vec B \nonumber \\
&=& - \frac{c}{4 \pi} \; \frac{\sin \theta}{r} \left[\frac{\partial^2 g(r,t)}{\partial r^2} 
- \frac{2 \, g(r,t)}{r^2} \right] {\hat \phi}
\eer

\subsection{Pure ohmic diffusion}
\ni In order to understand the ohmic dissipation of the field strength, let us consider equation
[\ref{emhd}] without the second term on the right hand side. The second term corresponds to the convective 
transport, which is relevant for material movement through the crust of an accreting neutron star.
Without this term equation [\ref{emhd}] takes the following form:
\beq
\frac{\partial \vec B}{\partial t} = 
- \frac{c^2}{4 \pi} \vec \nabla \times (\frac{1}{\sigma} \times \vec \nabla \times \vec B)
  \label{diffuse}
\eeq
For the uniform conductivity case, the above equation
takes the form of a pure diffusion equation (by virtue of 
the divergence-free condition for the magnetic fields): 
\beq
\frac{\partial \vec B}{\partial t} = 
 - \frac{c^2}{4 \pi \sigma} \nabla^2  \vec B
\eeq
The diffusion constant for the above equation is $c^2/4\pi\sigma$
and the time-scale characteristic of the process is:
\beq
\tau_{\rm diff} = \frac{4 \pi \sigma L^2}{c^2} \label{ediffconst}
\eeq
where $L$ is the length-scale associated with the underlying current distribution supporting the field. \\

\subsection{The field evolution equation}
\ni We use the form of $B(r, \theta, \phi)$ given by equation [\ref{eBrtheta}] 
to cast equation [\ref{diffuse}] in terms of the Stokes' 
stream function.\\
{I}. The left hand side:
\ber
\frac{\partial \vec B}{\partial t} 
&=& \frac{\partial \nabla \times \vec A}{\partial t} \nonumber \\
&=& \nabla \times \frac{\partial }{\partial t} \; \left[\frac{g(r, \theta) \; \sin \theta}{r}\right] 
\; {\hat \phi} \label{elhs}
\eer
{II}. The right hand side:
\[
\nabla \times \left[\frac{1}{\sigma} \nabla \times \vec B\right] =
\]
\[
\nabla \times \left[\frac{1}{\sigma} \nabla \times 
\left(\frac{2 \cos \theta \; g(r,t)}{r^2} \; {\hat r} 
- \frac{\sin \theta}{r}\; \frac{\partial g(r,t)}{\partial r} \; {\hat \theta}\right)\right] =
\] 
\beq
\nabla \times \left[ - \frac{1}{\sigma} \; \frac{\sin \theta}{r} \;
\left(\frac{\partial^2 g(r,t)}{\partial r^2} 
- \frac{2 g(r,t)}{r^2}\right) \; {\hat \phi}\right] \label{erhs2}
\eeq
Incorporation of the expressions [\ref{elhs}] and [\ref{erhs2}] in 
equation [\ref{diffuse}] then leads to:
\beq
\frac{\partial g(r,t)}{\partial t} =
 \frac{c^2}{4\pi \sigma} \left(\frac{\partial^2 g(r,t)}{\partial r^2} 
- \frac{2g(r,t)}{r^2} \right) \label{edgdt}
\eeq
The results presented in this paper are based on numerical solutions of 
the equation [\ref{edgdt}]. To solve this equation we used a modified
Crank-Nicholson scheme of differencing (Press et al. 1992) and
solved the relevant tri-diagonal matrices. We used
the standard boundary conditions, e.g. introduced by Geppert \& Urpin (1994).
We refer to Konar~(1997) for an extended  
discussion on the details of obtaining a stable mathematical solution,
and the physics of the boundary conditions.

{}


\begin{thebibliography}{}
  \bibitem{} Alpar M. A., Cheng A. F., Ruderman M. A., Shaham J., 1982,
             Nat. 300, 728
  \bibitem{} Baym G., Pethick C. J., Pines D., 1969, Nat 224, 673
  \bibitem{} Baym G., Pethick C. J., Sutherland P., 1971, ApJ 170, 299 
  \bibitem{} Beskin V. S., Gurevich A. V., Istomin Ya. N., 1988,
             Ap\&SS 146, 205
  \bibitem{} Bhattacharya D., 1989, in: X-ray Binaries,
             Proc. 23rd ESLAB Symp. vol.1, eds: \ J. Hunt, B. Battrick,
             ESA Noordwijk, The Netherlands
  \bibitem{} Bhattacharya D., Srinivasan G., 1995, in: X-ray Binaries,
             eds: W. H. G. Lewin, J. van Paradijs, E. P. J. van den Heuvel,
             Cambridge University Press
  \bibitem{} Blandford R. D., Applegate J. H., Hernquist L., 1983, MNRAS 204, 1025
  \bibitem{} Brown E. F., Bildsten L., 1998, ApJ 496, 915
  \bibitem{} Candy B. N., Blair D. G., 1983, MNRAS 281, 205
  \bibitem{} Candy B. N., Blair D. G., 1986, ApJ 307, 535
  \bibitem{} Geppert U., Urpin V., 1994, MNRAS 271, 490
  \bibitem{} Gil J. A., Han J. L., 1996, ApJ. 458, 265
  \bibitem{} Goldreich P., Julian W. H., 1969, ApJ 157, 869
  \bibitem{} Gould D. M., 1994, PhD thesis, Uni. of Manchester
  \bibitem{} Gudmundsson E. H., Pethick C. J., Epstein R. J., 1982, ApJ 259, L19
  \bibitem{} Gunn J. E., Ostriker J. P., 1970, ApJ 160, 979
  \bibitem{} Itoh N., Hayashi H., Yasuharu M., 1993, 418, 405
  \bibitem{} Jackson J. D., 1975, Classical Electrodynamics,
             2nd ed., John~Wiley~\&~Sons 
  \bibitem{} Johnston S., Galloway D., 1999, MNRAS 306, L50
  \bibitem{} Jones P. B., 1976, Ap\&SS 45, 369
  \bibitem{} Konar S., 1997, PhD thesis, Indian Institute of Science, Bangalore 
  \bibitem{} Konar S., Bhattacharya D., 1997, 284, 311
  \bibitem{} Konar S., Bhattacharya D., 1999, 303, 588
  \bibitem{} Lyne A. G., Manchester R. N., 1988, MNRAS 234, 477
  \bibitem{} Lyne A. G., Pritchard R. S., Graham-Smith F., Camilo F., 1996,
             Nat. 381, 497
  \bibitem{} Manchester R. N., Taylor J. H., 1977,
             Pulsars, Freeman, San Francisco
  \bibitem{} McKinnon M. M., 1993, ApJ. 413, 317
  \bibitem{} Negele J. W., Vautherin D., 1973, Nucl. Phys. A, 207, 298
  \bibitem{} Ostriker J. P., Gunn J. E., 1969, ApJ 157, 1395
  \bibitem{} Pacini F., 1967, Nat 216, 567
  \bibitem{} Pacini F., 1968, Nat 219, 145 
  \bibitem{} Pandey U. S., Prasad S. S., 1996, A\&A 308, 507
  \bibitem{} Press W. H., Teukolsky S. A., Vetterling W. T., Flannery B. P.,
             1992, Numerical Recipes in Fortran, Ch.~19, Cambridge University Press
  \bibitem{} Proszynski M., 1979, A\&A 79, 8
  \bibitem{} Rankin J. M., 1993, ApJ 405, 285
  \bibitem{} Ruderman M., Zhu T., Chen K., 1998, ApJ 492, 267
  \bibitem{} Sang Y., Chanmugam G., 1987, ApJ 323, L61 
  \bibitem{} Tauris T. M., Nicastro L., Johnston S., et~al.,~1994, ApJ 428, L53
  \bibitem{} Tauris T. M., Manchester R. N., 1998, MNRAS 298, 625 
  \bibitem{} Urpin V., Geppert U., Konenkov D., 1997, MNRAS 295, 907
  \bibitem{} Urpin V., Muslimov A. G., 1992, MNRAS 256, 261
  \bibitem{} van den Heuvel E. P. J., 1984, JA\&A 5, 209
  \bibitem{} van Riper K. A., 1991a, ApJS 75, 449
  \bibitem{} van Riper K. A., 1991b, ApJ 372, 251
  \bibitem{} Wiringa R. B., Fiks V., Fabrocini A., 1988, Phys. Rev. C, 38, 1010
  \bibitem{} Yakovlev D. G., Urpin V., 1980, SvA 24, 303

\end{thebibliography}
\end{document}